\documentclass[12pt]{iopart}
\usepackage{graphicx}

\newcommand{\ye}{Y_{e}}
\newcommand{\nuf}{\phi^{\mathrm{eff}}_{\nu}}
\newcommand{\anuf}{\phi^{\mathrm{eff}}_{\bar{\nu}}}

\begin{document}

\title{Production of $^{56}$Ni in black hole-neutron star merger accretion disk outflows}

\author{R. Surman$^{1}$, O.L. Caballero$^{2,3}$, G.C. McLaughlin$^{4}$, O. Just$^{5,6}$, H.-Th. Janka$^{5}$}

\address{$^1$Department of Physics and Astronomy, Union College, Schenectady, NY, USA}
\address{$^2$Joint Institute for Nuclear Astrophysics, Michigan State University, East Lansing, MI, USA}
\address{$^3$Institut f{\"u}r Kernphysik, Technische Universit{\"a}t Darmstadt, Darmstadt, Germany}
\address{$^4$Department of Physics, North Carolina State University, Raleigh, NC, USA}
\address{$^5$Max-Planck-Institut f{\"u}r Astrophysik, Garching, Germany}
\address{$^6$Max-Planck/Princeton Center for Plasma Physics (MPPC)}
\ead{surmanr@union.edu}

\begin{abstract}
The likely outcome of a compact object merger event is a central black hole surrounded by a rapidly 
accreting torus of debris.  This disk of debris is a rich source of element synthesis, the outcome 
of which is needed to predict electromagnetic counterparts of individual events and to understand 
the contribution of mergers to galactic chemical evolution.  Here we study disk outflow 
nucleosynthesis in the context of a two-dimensional, time-dependent black hole-neutron star merger 
accretion disk model.  We use two time snapshots from this model to examine the impact of the 
evolution of the neutrino fluxes from the disk on the element synthesis.  While the neutrino fluxes 
from the early-time disk snapshot appear to favor neutron-rich outflows, by the late-time snapshot 
the situation is reversed. As a result we find copious production of $^{56}$Ni in the outflows.
\end{abstract}

\submitto{\JPG}
\maketitle

\section{Introduction}

Two compact objects (black hole-neutron star or neutron star-neutron star) in a binary 
system will undergo orbital decay due to gravitational radiation emission and eventually 
merge.  The merger is thought to produce a larger, central compact object surrounded by a 
rapidly accreting torus of debris.  The debris torus may contribute to the energy needed to 
power a short-duration gamma-ray burst \cite{pac86,eic89,jan99,ros03}.  Some fraction of 
the mass of the system is expected to be ejected---promptly, in the tidal tails, and in 
outflows from the accretion disk.  Thus merger events are expected to be important 
contributors to galactic nucleosynthesis.  Understanding the nature of this contribution 
requires accurate predictions of the elemental composition of the ejected material.  Additionally, 
the decay of nuclear species produced in the merger will result in an observable 
electromagnetic counterpart \cite{met10,rob11,gor11}, which will be sensitive to the kinds 
and amounts of radioactive species produced \cite{bar13,gro13}.

The material ejected promptly or in the tidal tails is expected to be cold or only mildly 
heated and to retain much of the neutron richness of the original neutron star(s).  Thus it is 
an attractive site for the assembly of the heaviest elements in rapid neutron capture, or 
r-process, nucleosynthesis \cite{lat74,mey89,fre99b,gor11,kor12}.  The material that makes up 
the accretion disk outflows, on the other hand, will be highly processed before element 
synthesis occurs---first in the disk \cite{sur03}, where it is heated and the nuclei 
dissociated into their constituent nucleons, and then in the outflow \cite{sur04}, where the 
composition will be set by the weak interactions:
\begin{eqnarray} 
p + e^{-} &\rightleftharpoons & \nu_{e} + n\\ 
n + e^{+} &\rightleftharpoons & \bar{\nu}_{e} + p. 
\label{eq:weak}
\end{eqnarray}
As the accretion disk copiously emits neutrinos \cite{mcl06}, the above reverse reactions play 
a key role in determining the resulting element synthesis in the outflows.

Element synthesis in the hot outflows from merger-type black hole accretion disks has 
previously been addressed in, e.g., \cite{mcl05,sur06,met08,sur08,wan12,ros13}.  These found 
neutron-rich outflows to be possible in this environment, and thus suggested that merger-type 
disk winds may be an important source of particularly the lighter r-process nuclei.  In 
general neutron-rich outflows may result from two distinct mechanisms: the preservation of 
the high disk neutron-to-proton ratio in the outflow or the reset of the outflow 
neutron-to-proton ratio by the neutrino capture reactions from Eqns.\ 1,2.  The latter 
mechanism is the focus of this work.  For the neutrino interactions to drive the outflows 
neutron-rich, the antineutrinos must be sufficiently hotter than the neutrinos to overcome 
the neutron-proton mass difference, so that the rate of antineutrino captures on protons 
can exceed the neutrino capture rate on neutrons.  Calculations of disk neutrino decoupling 
surfaces, similar to protoneutron star (PNS) `neutrinospheres', show that as in the PNS the 
antineutrinos decouple deeper within the disk and are therefore hotter. However, the 
decoupling surfaces are not spherical as in the PNS but torus-shaped, so the antineutrinos 
emerge from a smaller region closer to the black hole.  This makes the antineutrinos 
subject to larger general relativistic effects \cite{cab09} and reduces the expected 
neutron richness of any neutrino-dominated outflows \cite{cab12}.

Here we address an additional consideration---does the disk neutrino emission persist long 
enough to make robustly neutron-rich outflows?  We consider the nucleosynthesis in the 
outflows from a time-dependent merger-type black hole accretion disk, carefully 
incorporating the important general relativistic effects on the neutrino emission from the 
disk.  We find that the mechanism for producing neutron-rich disk outflows via neutrino 
interactions does not operate in this model.  Instead, mildly proton-rich outflows are 
predicted.  We examine the resulting nucleosynthesis and show these outflows can result in 
significant $^{56}$Ni production.

\section{Black hole accretion disk model}

\begin{figure}
\centerline{\includegraphics[width=\textwidth]{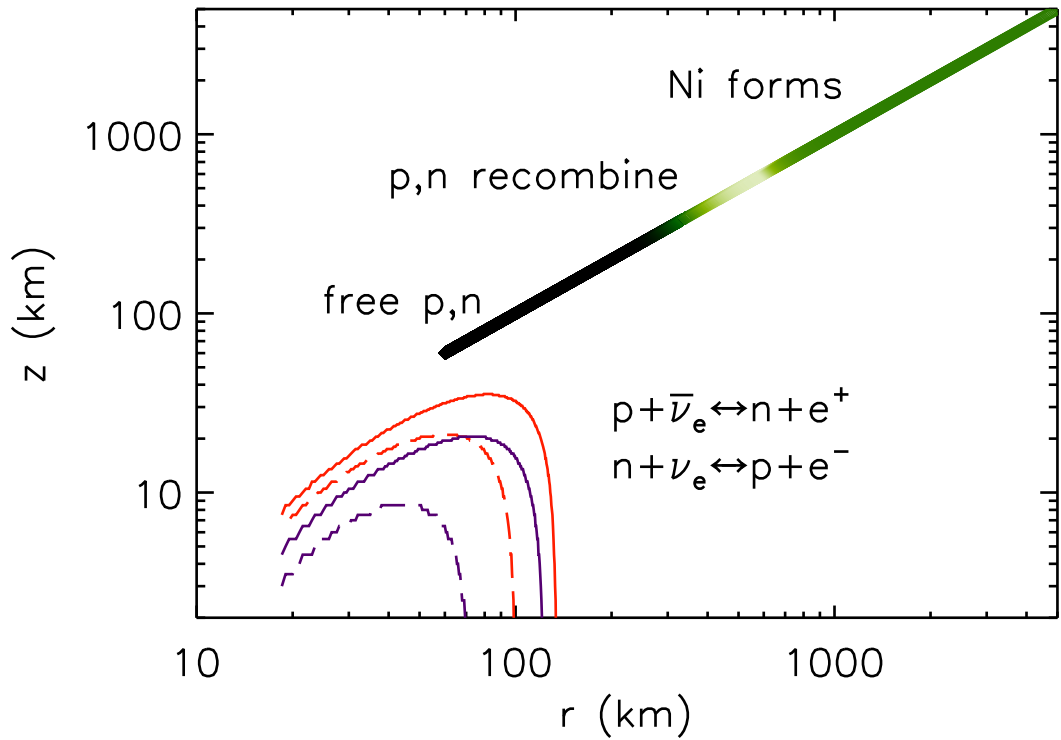}}
\caption{Neutrino (solid) and antineutrino (dashed) decoupling surfaces for the t=20 ms 
(thick red lines) and t=60 ms (thin purple lines) disk model snapshots.  Also shown is 
a sample radial outflow trajectory (shaded line), colored according to the mass fraction of 
alpha particles $X_{\alpha}$ along the trajectory, where black corresponds to 
$X_{\alpha}$=0 and white corresponds to $X_{\alpha}$=1.}
\label{fig:schem}
\end{figure}
We begin with a time-dependent model of a merger-type black hole accretion disk. 
The initial disk model is set up to be the equilibrium configuration of a 
constant angular momentum torus, similar to \cite{igu96}, where the black hole has a mass 
of 3 solar masses and spin parameter of $a=0.8$.  The simulated physics are Newtonian, 
except for the gravitational potential, which is given by the pseudo-relativistic 
Artemova-Novikov-Bjoernsson potential \cite{art96}.  The equation of state takes into account 
non-relativistic baryons (neutrons, protons, alphas, and one heavy nucleus), relativistic 
and/or degenerate electrons, and a plasma, and is similar to \cite{tim00}. At $t=0$ s the 
Shakura and Sunyaev alpha viscosity terms are turned on and the subsequent disk dynamics 
are followed.  

We take two time snapshots of this accretion disk model, at $t=20$ ms and $t=60$ ms, for careful study.  The first 
time snapshot is representative of the disk just after formation and corresponds to conditions similar to those 
from, for example, \cite{sur08}.  We consider outflows launched from this disk at $t=20$ ms.  As the outflowing 
material moves away from the disk, it will continue to be affected by the neutrino emission from the evolving disk.  
Thus, we use a second disk snapshot to capture the general behavior of the time evolution of the disk neutrino 
emission.  The neutrino interactions are the most important during the early part of the outflow, within the first 
tens of milliseconds from launch, so we choose $t=60$ ms for the second snapshot. For each disk snapshot, we 
calculate the electron neutrino and antineutrino decoupling surfaces as in \cite{cab09} and take the neutrino 
emission to be thermal, with a temperature equal to the local disk temperature at the point of decoupling.  The 
neutrino decoupling surfaces for the two disk snapshots are shown in Fig.~\ref{fig:schem}.  At early times, both 
the electron neutrino and antineutrino decoupling surfaces are large, and the antineutrinos emerge from deeper 
within the disk and are therefore hotter.  By $t=60$ ms, the disk has already lost enough mass to substantively 
change its structure---neutrino emission drops and the neutrino decoupling surfaces shrink.  These changes to the 
neutrino emitting regions will strongly impact the resulting nucleosynthesis in the outflows.

\section{Neutrino interactions in the outflow} 

To study the nucleosynthesis produced in hot outflows from this disk, we use an outflow model based 
on a streamline approach. We construct the model by first constraining the outflow to follow radial 
streamlines with zero vorticity. We use the steady flow approximation and consider the Bernoulli 
function along the streamline that defines the outflow trajectory.  In its simplest form, the 
Bernoulli function is
\begin{equation}
B=\frac{1}{2}v^{2}+h+\Phi,
\end{equation}
where $v$ is the outflow velocity, $h$ is the specific enthalpy, and $\Phi=-GM_{BH}/r$ is the 
gravitational potential with $M_{BH}$ the black hole mass, $r$ the distance to the black hole 
of a mass element, and $G$ the gravitational constant.  We write the specific enthalpy as
\begin{equation}
h=\frac{1}{\rho}\left[\sum_{i} n_{i}\mu_{i} + (kT)(s/k)n_{B}\right],
\end{equation}
where $\rho$ is the density, $T$ is the temperature, $n_{B}$ is the baryon number density, and $n_{i}$ and 
$\mu_{i}$ are the number density and chemical potential, respectively, of species $i$, where the sum over species 
includes electrons, positrons, neutrons, protons, alpha particles, and an average heavy nucleus\cite{lat91,mcl96}.  
We solve for the outflow hydrodynamics by further assuming a constant mass outflow rate, $\dot{M}=4\pi r^{2} \rho 
v$, which remains a free parameter in this formulation.  As in \cite{sur11}, we choose to follow adiabatic 
trajectories and take the Bernoulli function to be constant along the streamline.  The resulting trajectories, 
parameterized in entropy and mass outflow rate, are qualitatively similar to standard supernova neutrino-driven 
wind trajectories \cite{qia96}, parameterized in entropy and dynamic timescale $\tau$, where 
$r\sim r_{0}e^{t/\tau}$, so we adopt the latter prescription.

We launch material from the inner region of the disk, at $t=20$ ms and $r_{0}=30$ km; a schematic of the 
trajectory is given in Fig.~\ref{fig:schem}.  We take the starting composition of the 
outflow to be the local $t=20$ ms disk composition at that point, which is a mix of 
neutrons and protons with an electron fraction $\ye=1/(1+n/p)$ of 0.446.  We then evolve the 
composition, taking careful account of the weak reactions Eqns. 1,2 throughout.

\begin{figure}
\centerline{\includegraphics[width=\textwidth]{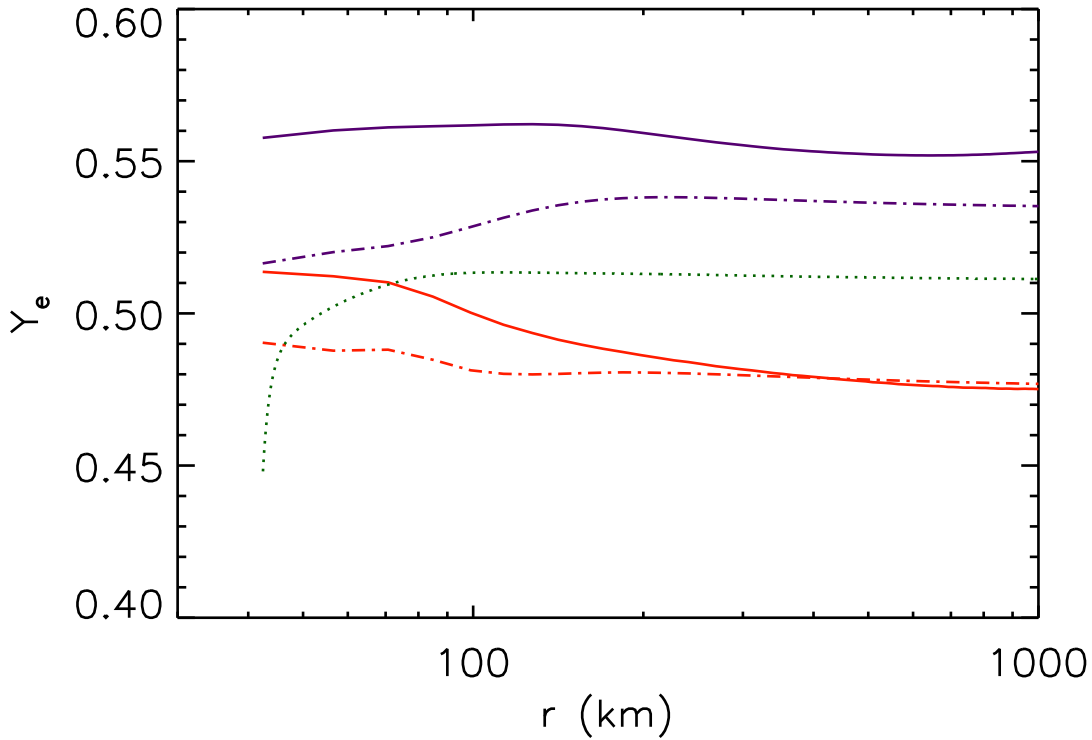}}
\caption{Neutrino-only equilibrium electron fractions $\ye^{\nu\mathrm{-only \, eq}}=1/(1+\lambda_{\bar{\nu}_{e}p}/\lambda_{\nu_{e}n})$ along a sample radial outflow trajectory for the $t=20$ ms 
(thick red lines) and the $t=60$ ms (thin purple lines) disk model snapshots, with (solid) and without (dot-dashed) 
GR effects on the neutrino fluxes.  The outflow electron fractions obtained in the full calculation, starting from the disk value of 0.446, tend to fall in 
between, as shown here for a $s/k=30$, $\tau=10$ ms outflow trajectory (dotted line).}
\label{fig:ye}
\end{figure}
Of particular importance are the neutrino capture reactions on nucleons, Eqns.\ 1,2. The 
rates of these reactions, $\lambda_{\nu_{e}n}$ and $\lambda_{\bar{\nu}_{e}p}$, 
respectively, are given by
\begin{eqnarray}
\lambda_{\nu_{e}n} & = & b \int^{\infty}_{0} \nuf
(E+\Delta)^{2} \sqrt{1-\left(\frac{m_{e}}{E+\Delta}\right)^2} W_{M} \, dE \\
\lambda_{\bar{\nu}_{e}p} & = & b \int^{\infty}_{\Delta+m_{e}c^{2}} \anuf 
(E-\Delta)^{2} \sqrt{1-\left(\frac{m_{e}}{E+\Delta}\right)^2} W_{\bar{M}} \, dE,
\end{eqnarray}
where $b=9.704\times10^{-50}$ cm$^{2}\cdot$keV$^{-2}$, $m_{e}$ is the electron mass, $\Delta$ is the neutron-proton 
mass difference, $W_{M} = 1 + 1.1(E/m_{n})$ and $W_{\bar{M}} = 1 - 7.1(E/m_{n})$ are the weak magnetism corrections 
\cite{hor02} and $\nuf$, $\anuf$ are the effective neutrino and antineutrino fluxes in units of 
1/(cm$^{2}\cdot$s$\cdot$keV).  The effective neutrino and antineutrino fluxes as observed by the outflowing fluid 
element above each disk snapshot are calculated as in \cite{cab12}.  We integrate the thermal fluxes described in 
Sec.\ 2 above over the solid angle of the emitting surface, both with and without general relativistic (GR) 
corrections to the neutrino energies and trajectories.  The neutrino and antineutrino capture rates are then 
calculated from each set of these fluxes along the radial outflow path.

To get a sense of what type of nucleosynthesis we can expect as well as the impact of the GR effects and the 
time-dependence of the neutrino fluxes, we examine the neutrino-only equilibrium electron fraction along the radial 
outflow path, calculated separately for the $t=20$ ms and $t=60$ ms disk snapshots.  The neutrino-only equilibrium 
electron fraction, $\ye^{\nu\mathrm{-only \, eq}}$, is the composition that would be expected if the neutrino 
interaction rates are much greater than the reverse weak rates of electron/positron capture, and is calculated by 
$\ye^{\nu\mathrm{-only \, eq}}=1/(1+\lambda_{\bar{\nu}_{e}p}/\lambda_{\nu_{e}n})$.  The results are shown in 
Fig.~\ref{fig:ye}.  We can see that if the $t=20$ ms disk snapshot were taken to be steady-state and the GR effects 
on the neutrino spectra are ignored, the outflows would tend to be driven neutron rich by neutrino interactions.  
However, once the GR effects are included, the situation changes, as first pointed out in \cite{cab12}.  The 
antineutrinos come from a smaller emitting surface closer to the black hole than the neutrinos, and so the 
antineutrino fluxes are more strongly shaped by the GR effects, particularly the redshifting of the neutrino 
energies.  Thus at the start of the trajectory the strong neutron-richness predicted in the no-GR 
$\ye^{\nu\mathrm{-only \, eq}}$ is lost.  By $t=60$ ms, the antineutrino emitting surface has become so small that, 
even though the antineutrinos are still hotter than the neutrinos, the material along the outflow trajectory `sees' 
more neutrinos than antineutrinos, and $\lambda_{\nu_{e}n}>\lambda_{\bar{\nu}_{e}p}$ and $\ye^{\nu\mathrm{-only \, 
eq}}>0.5$.

The effects of the time dependence of the neutrino fluxes are estimated by a simple linear 
scheme.  The outflow is launched at $t=20$ ms, and from $t=20$ ms to $t=60$ ms the neutrino 
capture rates are taken to be
\begin{equation}
\lambda(t) = \lambda^{t=20 \, \mathrm{ms}} + \left(\frac{t-20 \, \mathrm{ms}}{40 \, \mathrm{ms}}\right) \left(\lambda^{t=60 \, \mathrm{ms}} - \lambda^{t=20 \, \mathrm{ms}}\right)
\end{equation}
and for $t>60$ ms $\lambda(t) = \lambda^{t=60 \, \mathrm{ms}}$, where $\lambda^{t=20 \, \mathrm{ms}}$ and 
$\lambda^{t=60 \, \mathrm{ms}}$ are the interaction rates calculated with the $t=20$ ms and $t=60$ ms disk neutrino 
fluxes, respectively.  This choice of an interpolation scheme captures the general features of the time dependence 
of the neutrino fluxes and should influence only the details of the predicted nucleosynthetic yields and not the 
general conclusions.

\section{Element synthesis}

Using the neutrino interaction rates described above, we follow the element synthesis as in \cite{sur11}, first 
with an NSE calculation \cite{mcl96} and then switching over to a full nuclear network \cite{hix99} at $T=10$ GK.  
We examine a range of our outflow parameter space: entropies $20 \leq s/k \leq 80$ and effective dynamic timescales 
10 ms$\leq \tau \leq100$ ms. As anticipated, the actual evolution of the electron fraction in the outflow tends to 
lie somewhere in between the $\ye^{\nu\mathrm{-only \, eq}}$ values for the two disk snapshots, as shown for an 
example full trajectory in Fig.~\ref{fig:ye}.  As a result, we find proton-rich nucleosynthesis products in almost 
all of the outflows studied.  

\begin{figure}
\centerline{\includegraphics[width=\textwidth]{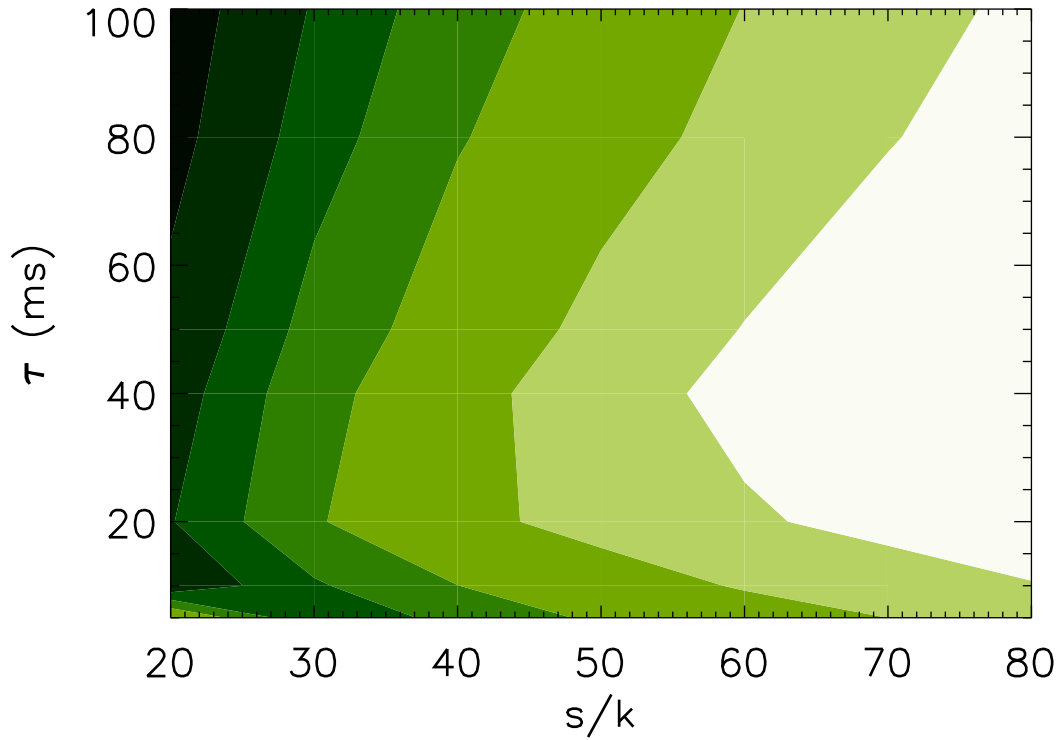}}
\caption{Nickel-56 production for the outflow parameter space---entropy per baryon $s/k$ and 
effective dynamic timescale $\tau$---sampled in this work.  The shaded regions correspond to nickel 
mass fractions of 0.5, 0.4, 0.3, 0.2, 0.1, and less than 0.1, from darkest to lightest.}
\label{fig:ni}
\end{figure}
The nucleus with $A>4$ most vigorously produced is $^{56}$Ni. The final mass fractions of 
nickel in the outflows are shown in Fig.~\ref{fig:ni}.  For mildly heated outflow material 
over half is ejected as $^{56}$Ni.  Increases in heating tend to favor lighter nuclei, and 
so for trajectories with higher entropy most of the material is left as alpha particles.  
Still, $^{56}$Ni remains the most abundantly produced $A>4$ nucleus.

Note that the prediction of nickel-rich outflows for this disk model depends strongly on the neutrino physics and 
is largely insensitive to the particular outflow parameterization chosen.  We draw the same conclusions regarding 
$^{56}$Ni production when we repeat the nucleosynthesis calculations using the outflow model from \cite{sur06}.

\section{Conclusion}

Merger-type black hole accretion disks have been shown to be attractive potential sites for 
neutron-rich nucleosynthesis.  Here we examine one potential mechanism for producing 
neutron-rich outflows---via neutrino interactions in the outflow---and note that its 
operation is not ubiquitous.  In our calculations, we find that the combination of general 
relativistic effects on the neutrino fluxes as well as the rapid evolution of the neutrino 
emission surfaces have the opposite effect.  The resulting proton-rich outflows will not 
produce r-process nuclei but may be a rich source of $^{56}$Ni.

Any $^{56}$Ni produced will decay first into $^{56}$Co then $^{56}$Fe.  Such decays power 
the light curves of core-collapse supernovae and so could potentially produce an observable 
electromagentic counterpart to the merger event.  If $\sim 1\%$ of the disk mass is ejected 
in winds that are driven proton-rich by the neutrino physics, on the order of $10^{-3} 
M_\odot$ of $^{56}$Ni may be ejected.  This is sufficient to generate a bright optical peak 
\cite{bar13}, which would be in addition to the infrared peak expected to be due to the 
radioactive r-process ejecta \cite{met10,rob11,gor11}.

\section*{Acknowledgments}
This work was supported by the Department of Energy under contracts DE-FG02-05ER41398 (RS) and 
DE-FG02-02ER41216 (GCM).  OJ and TJ acknowledge support by the Computational Center for
Particle and Astrophysics (C2PAP) as part of the Cluster of Excellence
EXC 153 ``Origin and Structure of the Universe" and by the Max-Planck-Princeton Center for Plasma Physics.

\end{document}